\documentclass[twocolumn,prl,showpacs]{revtex4}

\usepackage{graphicx}
\usepackage{bm}

\begin{document}

\title{How the hydrogen bond in NH$_4$F is revealed with Compton scattering}

\author{B. Barbiellini$^{1}$, Ch. Bellin$^{2}$, G. Loupias$^{2}$,
T. Buslaps$^{3}$ and A. Shukla$^{2}$}
\affiliation{$^{1}$ Department of Physics, Northeastern University,
Boston, MA 02115, USA\\
$^{2}$ 
Institut de Min\'eralogie et Physique de la Mati\`ere Condens\'ee, 
Universit\'e Pierre et Marie Curie, Campus Boucicaut, 140 rue de Lourmel, 75015 
Paris, France\\
$^{3}$European Synchrotron Radiation Facility, BP220 Grenoble,
France}

\date{\today}



\begin{abstract}
In order to probe electron wave functions
involved in the bonding of NH$_4$F,
we have performed Compton scattering experiments in an
oriented single crystal and in a powder.
{\em Ab initio} calculations of the Compton profiles
for NH$_4$F and NH$_4$Cl
are used to enlighten the nature of the bonds in the 
NH$_4$F crystal.
As a consequence, we are able to show
significant charge transfer in the ammonium ion
which is not observable using other methods. 
Our study provides a compelling proof for 
hydrogen bond formation in NH$_4$F.
\end{abstract}

\maketitle

The distinction between ionic and
hydrogen bonding is not always 
clear \cite{hughes}.
Ammonium salts, in general are
similar to the potassium and
rubidium salts in crystal form
and in other physical properties 
\cite{pauling} and thus
are clearly ionic in nature.
This similarity is due to the size of the ammonium ion,
NH$_4^+$, which is almost equal to the size
of K$^+$ and Rb$^+$ ions.
Remarkably, NH$_4$F, instead of having
a CsCl or NaCl structure
like other ammonium salts which are ionic in nature, is
isostructural to ice Ih with NH$_4^+$
and F$^-$ ions alternating in the oxygen positions.
Besides, ammonium fluoride and ice can form mixed
crystals \cite{nature54}.
The fact that ammonium fluoride deviates from the
standard isomorphism and
prefers to adopt the ice structure
is ascribed to the presence of significant
hydrogen bonding in this compound \cite{mair,brown}.

While ice has been the subject of many studies,
NH$_4$F has not been as widely studied.
The crystal structure has been determined
by Adrian and Feil \cite{adrian},
the electrostatic properties have been studied by van Beek
{\em et al.} \cite{beek}, {\em Ab initio} calculations
have been carried out by Alavi {\em et al.} \cite{alavi}
and by van Reeuwijk {\em et al.} \cite{van}.
Given the ionic nature of ammonium salts, it is
tempting to consider
the force between the NH$_4^+$
and F$^-$ ions to be essentially
an electrostatic interaction.
Nevertheless, charge density distribution does not provide a good 
criterion  for establishing  the existence of hydrogen bonds, as shown 
for example by Alavi {\em et al.} \cite{alavi} for this compound.
In this work we investigate the nature of chemical bonding in
NH$_4$F and show, using the study of electron momentum distributions, 
that we detect the presence of hydrogen bonding in ammonium fluoride.

Recently, Compton scattering, which is inelastic X-ray
scattering at large energy and momentum transfers 
\cite{Compton_book,kaplan}
has provided fundamental information on the quantum nature 
of the hydrogen bond in ice and water 
\cite{prl_ice,ghanty,romero,ragot,prb_bba02,prb_sit,prl_nygard}.
The measured Compton profile (CP), $J(p_z)$
represents the double integral of the ground-state
total electron momentum distribution (EMD) 
$\rho ({\it\mbox{\boldmath$p$}})$:
\begin{equation}
\label{eq_cp}
J(p_z) = \int\!\!\!\int\rho ({\bf p})dp_x dp_y,
\label{eq1} 
\end{equation}
where $p_z$ lies along the scattering vector of the x-rays.
Our Compton scattering experiments on 
NH$_4$F will show how electronic states are modified
by the hydrogen bond in contast to the
ionic scenario of NH$_4$Cl.

%
%
The CPs of the crystalline $a$,
$c$ directions
and the spherical average of a powder
were measured at
the European Synchrotron Radiation Facility
in Grenoble on high energy
beamline (ID15B),
using the scanning Compton spectrometer\cite{suortti}.
The energy of incident photons was set to 29.74 keV
by use of a Si(111)
cylindrically bent crystal and the scattering angle was 173$^\circ$.
The collected Compton spectra were energy analyzed using the Si(400)
analyzer and detected by a NaI scintillator counter.
The energy-dependent resolution function is deduced from the full width
at half maximum (FWHM) at the thermal diffuse scattering peak which was
0.16 (a.u.) which provides at the Compton
peak 0.13 a.u. of
resolution. After subtracting the background,
energy dependent corrections have been
performed:
absorption in analyzer and detector, detector efficiency,
analyzer reflectivity \cite{balibar}.
After normalization to the number of electrons, the contribution due
to multiple scattering  was subtracted from the total measured profile
\cite{chomilier}.
In order to obtain the valence electrons CPs,
a quasi-self-consistent-field calculated core CP
is then subtracted from the total corrected measured profile
\cite{issolah}.

\begin{figure}
\begin{center}
\includegraphics[width=\hsize]{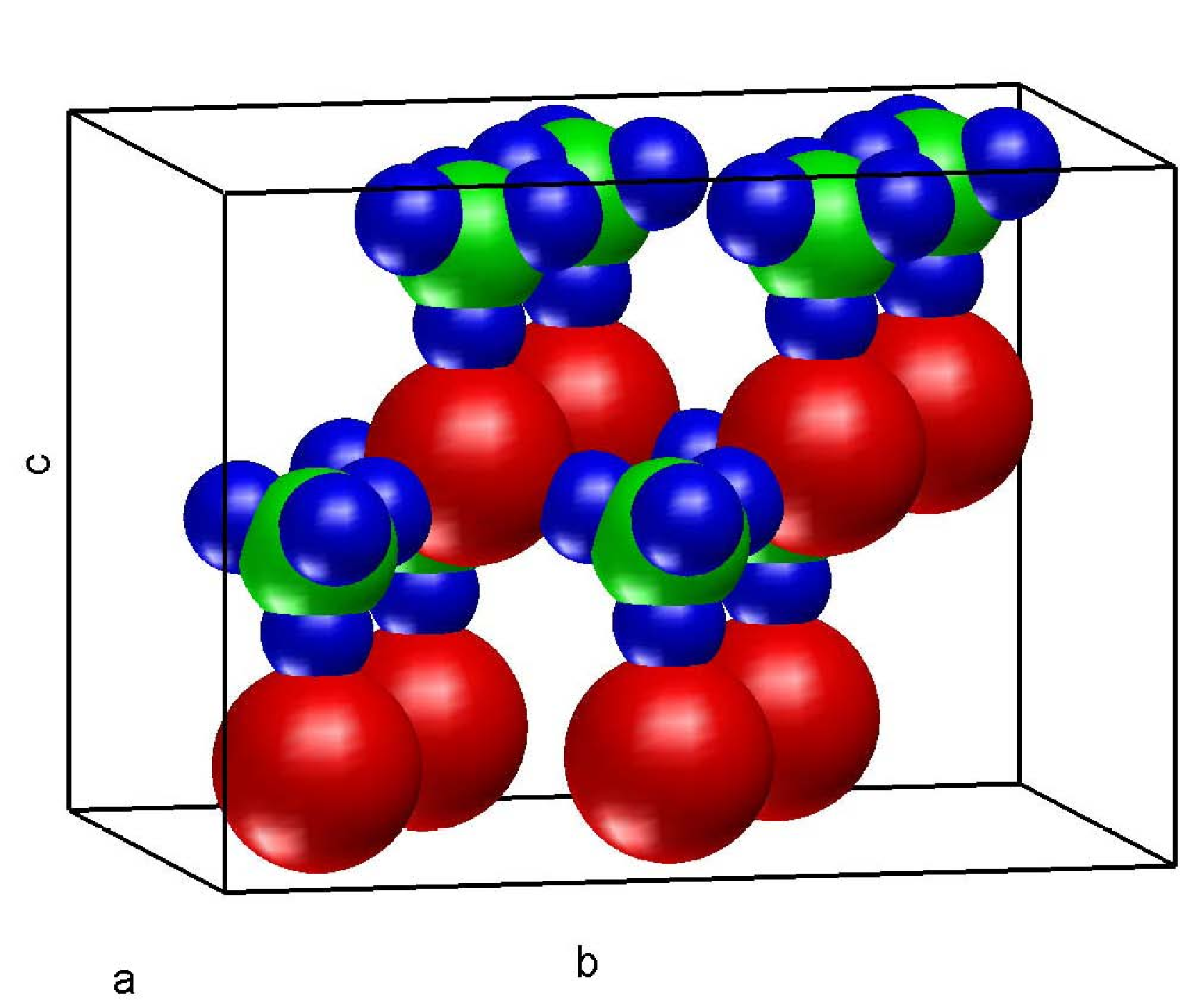}
\end{center}
\caption{(Color on-line) 48 atom cluster 
extracted from the NH$_4$F crystal.
The (red) largest spheres are F atoms, 
the small (blue) spheres are H atoms and the 
medium (green) spheres are N atoms.
The crystallographic directions
$a$, $b$ and $c$ are shown along the 
edges of a box containing the cluster.}
\label{fig_struct}
\end{figure}

The program used for our
{\em ab initio} calculations was
CRYSTAL98 \cite{crystal98},
which is especially suitable for
molecular crystals.
We have used the
hexagonal structure of the NH$_4$F crystal
at room temperature with point group
P63mc, $a=4.37$ \AA\ , $c=7.17$ \AA\
and the experimental
atomic positions \cite{beek}.
This crystal structure is illustrated 
in Fig.~\ref{fig_struct}.
The occupied orbitals used to determine
the EMD and the CPs were calculated
using a restricted Hartree-Fock scheme 
\cite{prb_bba02}.
%
%
%

%
%
\begin{figure}
\begin{center}
\includegraphics[width=\hsize]{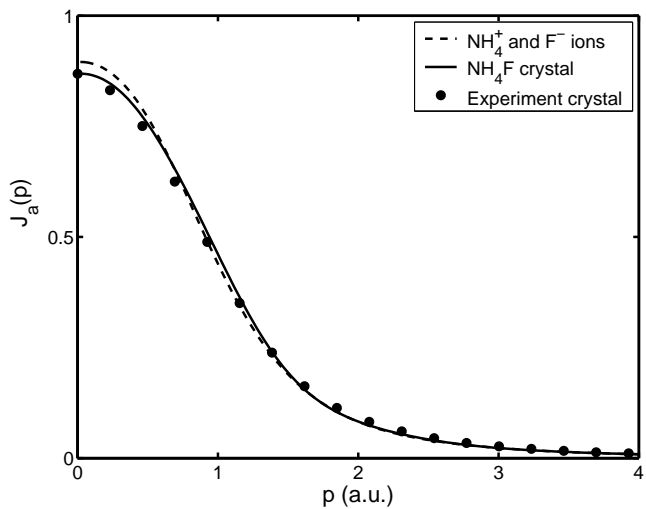}
\end{center}
\caption{
Comparison of the theoretical and experimental
[100] valence CPs.
The dots are the experimental points.
The dashed line is for
the superposition
of isolated NH$_4^+$
and F$^-$ ions.
The solid line is for 
the NH$_4$F crystal.
The curves are normalized to unit area.
Symbol size is representative of error bars.}
\label{fig1}
\end{figure}
In Fig.~\ref{fig1}, we show two computed 
valence CPs for NH$_4$F
(convoluted 
with experimental resolution and normalized 
to unit area)
together with the experimental data
along the crystallographic direction $a$.
In the first calculation, represented by a
dashed line,
we have built the CP as a superposition
of CPs of isolated ions NH$_4^+$
and F$^-$ arranged in the same geometry as
in the crystal.
In comparison with the 
experiment, the profile of the isolated ions 
is found to be higher due 
to the absence of bonding \cite{prb_sit}.
In the second calculation, represented by a
solid line, the two ionic fragments
can interact, and an excellent
agreement between
the crystal calculation
and the experiment is obtained.

%
%
\begin{figure}
\begin{center}
\includegraphics[width=\hsize]{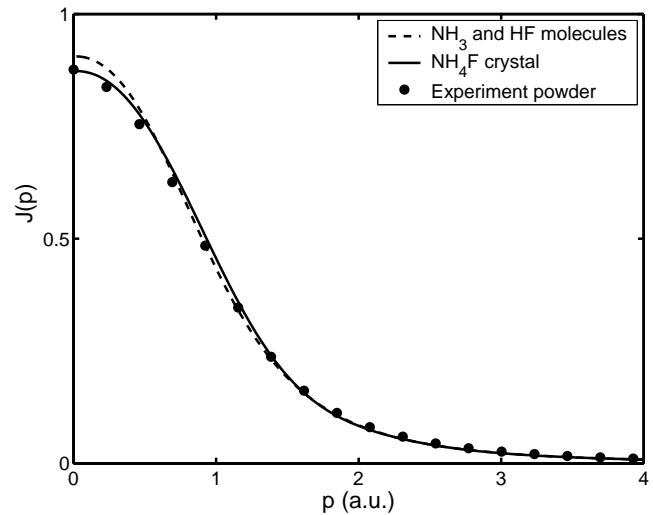}
\end{center}
\caption{Comparison of the theoretical
spherical averaged valence CPs
and the experiment for the powder.
The dots are the experimental points.
The dashed line is for
the superposition
of neutral NH$_3$
and HF molecules.
The solid line is for the NH$_4$F crystal.
The curves are normalized to unit area.
Symbol size is representative of error bars.}
\label{fig2}
\end{figure}
Likewise, in Fig.~\ref{fig2}, we show that the
spherical averaged CP
for the NH$_4$F crystal calculation (solid line)
agrees remarkably well with
the experimental CP of the powder
in contrast to a calculation
for two non-interacting neutral molecules
NH$_3$ and HF (dashed line).
In this molecular computation,
the intramolecular
atomic distances
are the same
as the corresponding distances in the crystal.
Interestingly, the spherical
averaged CP for two
non-interacting
neutral molecules NH$_3$ and HF
turns out to be very
similar to the spherical averaged CP
for the isolated ions
NH$_4^+$ and F$^-$.
In reality, the main 
difference between the ions and the molecules
cases is an exchange of a proton since 
the HF molecule can be formed by the capture
of a proton by the F$^-$ from the ammonium ion.  
However, the X-ray Compton scattering has not 
direct access to the proton wavefunction.

\begin{figure}
\begin{center}
\includegraphics[width=\hsize]{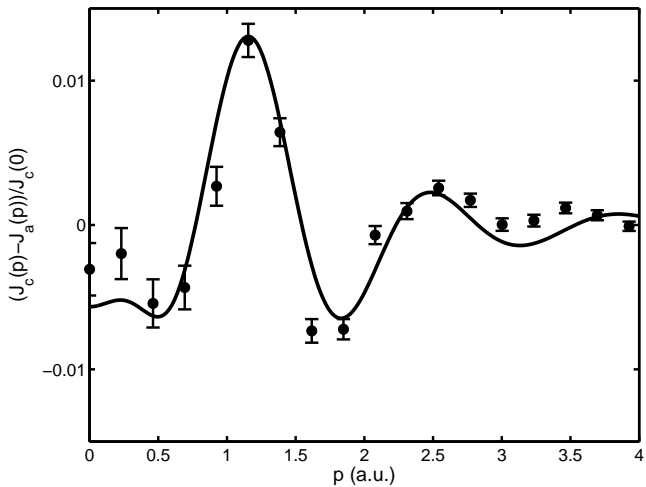}
\end{center}
\caption{
Comparison of the theoretical and experimental
Compton anisotropy [001] - [100].
The circles are the experimental points while
the solid line is for the {\em ab initio} simulation
of the NH$_4$F crystal.
The amplitude of the experimental anisotropy 
has been rescaled as in Ref.~\cite{prl_ice}.}
\label{fig_anis}
\end{figure}

In order to study bonding effects,
we now discuss the CP anisotropy
\cite{prl_ice}, that is, the difference
between measured Compton profiles along two
crystalline directions.
In the present case, we consider
\begin{equation}
A(p_z) =J_c(p_z)-J_a(p_z),
\label{eq2}
\end{equation}
where $J_c(p)$ is the CP along the crystallographic direction $c$
which is also the direction along which hydrogen bonds are aligned,
and $J_a(p)$ is the CP along the crystallographic direction $a$.
The anisotropy $A(p_z)$ has the
advantage of highlighting the changes introduced by the bonding.
The shape of the anisotropy is very similar to that found 
in ice $A(p_z)$  \cite{prl_ice} as shown in Fig.~\ref{fig_anis}.

%
%

\begin{figure} 
\begin{center} 
\includegraphics[width=\hsize]{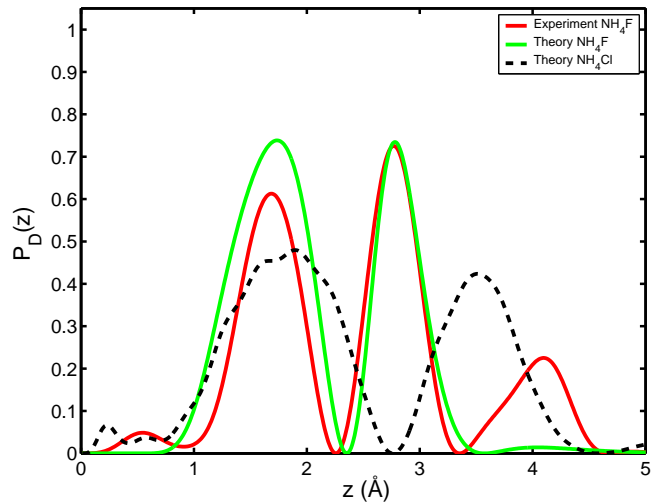} 
\end{center} 
\caption{(Color online) $P_D(z)$ or the power 
spectrum of the Compton anisotropy 
for NH$_4$F (experiment and theory) 
and NH$_4$Cl (theory). 
The curves are normalized to unit area.} 
\label{fig3} 
\end{figure}

The effects produced 
by the hydrogen bond on 
the electronic wavefunctions can be 
extracted by studying the power density \cite{prl_ice,PD}
\begin{equation} 
P_D(z)= \left|\int dp_z A(p_z) \exp(ip_{z}z)\right|^2. 
\label{eq3} 
\end{equation} 
The peaks in the power density $P_D(z)$ 
indicate characteristic 
distances over which wave functions are coherent 
in given crystallographic directions. 
Figure~\ref{fig3} shows that the 
experimental data and the theory are in good
agreement concerning the locations
of the two main peaks.
Nevertheless, the long range peak at about 
4 $\mbox{\AA}$  in the
experimental anisotropy is largely 
reduced in the calculation.
A cause of this discrepancy
could be the Hartree-Fock 
tendency to overestimate 
wavefunction localization.
However, the CP in momentum 
space is not too sensitive 
to this overestimation.

In a true ionic crystal the main differences from the isotropic
picture occur in the region where the ionic spheres are in contact
as in the hard sphere model. Thus, in the picture developed by Pattison 
{\em et al.} \cite{pattison} one expects peaks at distance
corresponding to the ionic diameters. The first main peak of $P_D(z)$
should correspond to the cation diameter while the second main peak 
should give the anion diameter. 
Figure~\ref{fig3} also compares NH$_4$F 
to the more {\em ionic} 
salt NH$_4$Cl \cite{nh4cl}.
One can see that the second main peaks
for NH$_4$F and NH$_4$Cl \cite{nh4cl}
are close to accepted values for F$^-$ and Cl$^-$ diameters
(2.7 $\mbox{\AA}$ and 3.6 $\mbox{\AA}$ respectively, \cite{ionrad}).
The first peak is however at a distance systematically smaller than
that of the hard sphere ionic diameter of ammonium (2.8 $\mbox{\AA}$)
even in NH$_4$Cl signifying that the purely ionic, hard sphere model is 
not adequate.
Moreover, the distance corresponding to this peak
is smaller for both the theoretical calculation and the experiment
in NH$_4$F than the calculation for NH$_4$Cl though the ion is identical.
As for NH$_4$F, the interpretation becomes straightforward when the 
connection is made with results obtained for the hydrogen bonded 
ice Ih since the measured and calculated anisotropy bear a striking 
resemblance. In the more appropriate hydrogen bond picture the two 
predominant peaks of $P_D(z)$ located at about 1.7 $\mbox{\AA}$ and 
2.75 $\mbox{\AA}$ can be assigned to the hydrogen bond length F-H (1.67 
$\mbox{\AA}$) and the nearest neighbor F-N distance (2.77 
$\mbox{\AA}$), respectively implying the existence of coherent charge 
transfer as in the case of ice Ih \cite{prb_bba02}.

%
%

In conclusion, our study shows that high-resolution directional 
CPs of ammonium fluoride provide clear signatures of the hydrogen bond.
The ionic picture is just a starting point since highly charged states 
such as the NH$_4^+$ and F$^-$ are mitigated by the hydrogen bond:
bonding electrons are shared in states that
cannot be assigned solely to ammonium or F.

This work was supported by the US Department of Energy, Office of Science, 
Basic Energy Sciences contract DE-FG02-07ER46352, and benefited from the
allocation of computer time at NERSC and at Northeastern
University Advanced Scientific Computation Center (NU-ASCC).


\begin{thebibliography}{99}
\bibitem{nh4cl}
For the NH$_4$Cl crystal calculations, we have used the
cubic ordered phase with space group P$\bar{\mbox{4}}$3m
and lattice parameter $a=3.82$ \AA\ \cite{alavi_nh4cl}.
\end{thebibliography}
\end{document}